\documentclass[conference, twocolumn, 9pt]{IEEEtran}	

\PassOptionsToPackage{utf8}{inputenc}
\usepackage{inputenc}
\usepackage{cite}
\PassOptionsToPackage{T1}{fontenc} 
\usepackage{fontenc}
\PassOptionsToPackage{dvipsnames}{xcolor}
\usepackage{soul} 

\PassOptionsToPackage{fleqn}{amsmath}       
\usepackage{amsmath}
\usepackage{mathtools}
\usepackage{xfrac}
\usepackage{siunitx}
\usepackage[inline]{enumitem}

\usepackage[vlined,ruled,linesnumbered,longend]{algorithm2e}
\usepackage{graphicx}
\usepackage{textcomp}
\usepackage{xspace}
\usepackage[size=tiny]{todonotes}
\usepackage[hidelinks]{hyperref}

\usepackage{subfig}
\usepackage{multirow}
\usepackage{multicol}
\usepackage[export]{adjustbox}
\usepackage{float}
\usepackage{booktabs}
\usepackage[super]{nth}

\usepackage{bm} 

\newcommand{\ie}{i.\,e.\xspace}

\newcommand{\eg}{e.\,g.\xspace}

\newcommand{\etal}{\textit{et~al}.\xspace}

\newcommand{\siqad}{\emph{SiQAD}\xspace}

\newcommand{\refsec}[1]{Section~\ref{sec:#1}}
\newcommand{\reffig}[1]{Fig.~\ref{fig:#1}}

\newcommand{\reftab}[1]{Table~\ref{tab:#1}}

\newcommand{\refalgo}[1]{Alg.~\ref{algo:#1}}
\newcommand{\refline}[1]{Line~\ref{line:#1}}
\newcommand{\reflines}[2]{Lines~\ref{line:#1}--\ref{line:#2}}
\newcommand{\refequa}[1]{Eq.~(\ref{eq:#1})}

\makeatletter
\IEEEtriggercmd{\reset@font\normalfont\fontsize{7.2pt}{8.10pt}\selectfont}
\makeatother
\IEEEtriggeratref{1}

\begin{document}
	
	\title{Near-Optimal physical Simulation of Silicon Dangling Bond Logic} 
	\title{A Simple and Effective Simulation of Silicon Dangling Bond Logic} 
	\title{Overcoming the Time-to-Solution Trade-off in physical Simulation of Silicon Dangling Bond Logic} 
	\title{Overcoming the Trade-off between Runtime and Quality \\ in Physical Simulation of Silicon Dangling Bond Logic} 
	\title{\emph{QuickSim:} Efficient \emph{and} Accurate \linebreak Physical Simulation of Silicon Dangling Bond Logic \vspace{-0.2cm}} 

	\author{\IEEEauthorblockN{Jan Drewniok\IEEEauthorrefmark{1}, Marcel Walter\IEEEauthorrefmark{1}, Samuel Sze Hang Ng\IEEEauthorrefmark{2}, Konrad Walus\IEEEauthorrefmark{2}, and Robert Wille\IEEEauthorrefmark{1}\IEEEauthorrefmark{3}} \\[-1ex]
		\IEEEauthorblockA{\IEEEauthorrefmark{1}Chair for Design Automation, Technical University of Munich, Germany \\
			Email: \{jan.drewniok, marcel.walter, robert.wille\}@tum.de} \\[-2ex]
		\IEEEauthorblockA{\IEEEauthorrefmark{2}Department of Electrical and Computer Engineering, University of British Columbia, Vancouver, Canada \\ %
			Email: \{samueln, konradw\}@ece.ubc.ca} \\[-2ex]
		\IEEEauthorblockA{\IEEEauthorrefmark{3}Software Competence Center Hagenberg GmbH, Austria}}
	\maketitle
	\IEEEpeerreviewmaketitle
	
	\begin{abstract}
		Silicon Dangling Bonds have established themselves as a promising competitor in the field of \mbox{beyond-CMOS} technologies. Their integration density and potential for energy dissipation advantages of several orders of magnitude over conventional circuit technologies sparked the interest of academia and industry alike. While fabrication capabilities advance rapidly and first design automation methodologies have been proposed, physical simulation effectiveness has yet to keep pace. Established algorithms in this domain either suffer from exponential runtime behavior or subpar accuracy levels.		
		In this work, we propose a novel algorithm for the physical simulation of Silicon Dangling Bond systems based on statistical methods that offers both a \mbox{time-to-solution} and an accuracy advantage over the state of the art 
		by more than one order of magnitude and a factor of more than three, respectively, as demonstrated by an exhaustive experimental evaluation.

		%
	\end{abstract}
	
	
	\section{Introduction \& Motivation}

	
	
	Recent years have seen an increase in scientific interest in the \emph{Silicon Dangling Bond}~(SiDB) logic platform---an emerging computational \mbox{beyond-CMOS} nanotechnology~\cite{D0NR08295C, wyrick2019atom, haider2009controlled, huff2017atomic, pavlicek2017tip, achal2018lithography, Wolkow14}. Of particular significance are its \mbox{sub-nanometer} elementary devices that allow for an integration density improvement of several orders of magnitude over current CMOS fabrication nodes~\cite{huff2018binary, huff2017atomic, achal2018lithography, wang2020atomic, haider2009controlled, Wolkow14, pitters2011charge, rashidi2018initiating}; and its properties to compute Boolean logic without the flow of electrostatic current but, instead, via the Coulombic repulsion of charges~\cite{huff2018binary, huff2017atomic}. By this, SiDB logic promises \mbox{ultra-low} energy dissipation and establishes itself as a \mbox{highly-anticipated} green competitor in the \mbox{beyond-CMOS} \mbox{domain \cite{landauer1961irreversibility, keyes1970minimal, lent2006bennett, toth1999quasiadiabatic, huff2018binary, rashidi2018initiating, haider2009controlled}}. Moreover, it has been proposed as a candidate for the integration of quantum computers~\cite{dzurak2001construction}, and---since SiDBs are fabricated on silicon as well---with conventional CMOS circuitry~\cite{Wolkow14}. 
	
	
	Motivated by this, the scientific community already proposed gate and circuit libraries~\cite{ng2020siqad, vieira2022three, walter2022hexagons, chiu2020poissolver, ng2020thes, chiu2020thes, bahar2020atomic} as well as design automation solutions~\cite{walter2022hexagons, lupoiu2022automated, ng2020siqad} for SiDB logic. But also commercially, this technology gains more and more interest as confirmed, \eg, by the recently founded \emph{Quantum Silicon Inc.} which 
	aspires to be among the first industry adopters of SiDBs and already secured \mbox{multi-million} dollar investments for this purpose~\cite{Wolkow2021, Wolkow2021a}.
	
	However, the rapid advancement of SiDB fabrication capabilities that goes along with these developments~\cite{haider2009controlled, huff2017atomic, pavlicek2017tip, achal2018lithography} puts pressure on both design and simulation tools to keep pace. Particularly physical simulation, the foundation of system validation, poses to be challenging. To reliably predict the behavior of any given system of~$n$ SiDBs, a high-dimensional optimization problem must be solved to capture all physically relevant effects. In its essence,~$3^n$ charge configurations have to be enumerated, checked for physical feasibility, and their individual energies simulated such that the lowest of those can be picked
	---constituting a highly non-trivial (in fact, exponential) problem. 
	

	Thus far, only two 
	approaches 
	addressing this problem have been proposed in the literature: \emph{ExhaustiveGS}~(ExGS)~\cite{vieira2022three, ng2020thes}, which performs an exhaustive search of the entire (exponential) search space, and \emph{SimAnneal}~\cite{ng2020siqad}, 
	which employs probabilistic sampling and, hence, offers a heuristic approach that trades runtime for completeness.
	Accordingly, ExGS is severely limited in its runtime efficiency, while SimAnneal only provides an approximation and, hence, is not guaranteed to 
	determine an optimal (\ie, perfectly accurate) solution. 
	
	In this paper, we propose a novel algorithm (\emph{QuickSim}) that aims to be both efficient \emph{and} accurate. To this end, we are making use of effective search space pruning by incorporating ideas from statistical methods like \mbox{\emph{max-min diversity distributions}~\cite{resende2010grasp}}.
	
	Exhaustive experimental evaluations (covering the simulation of common logic operations, established gate libraries, as well as random instances to demonstrate the performance on future layouts) confirm that, the resulting approach leads to 
	a \mbox{time-to-solution} advantage of more than one order of magnitude. At the same time, it outperforms SimAnneal in terms of solution accuracy by more than a factor of four on \mbox{randomly-generated} SiDB layouts (which are known to be the most challenging use case) and by roughly a factor of three on established gate layouts. 

	The remainder of this work is structured as follows: in an effort to establish this paper as a \mbox{self-contained} article, \refsec{prelims} reviews preliminaries on SiDB systems. 
	\refsec{simulations:approaches} discusses existing simulation approaches. \refsec{proposed} introduces \emph{QuickSim} in detail. \refsec{eval} comprises of a case study showcasing the benefits of \emph{QuickSim} over the state of the art. Finally, \refsec{concl} concludes the paper.
	
	
	\section{Preliminaries} \label{sec:prelims}
	
	This section first provides an overview of the SiDB logic platform. Afterward, the physical simulation of SiDBs is elaborated on. 
	
	\subsection{The SiDB Logic Platform} \label{sec:prelims:sidbs}
	
\begin{figure}[t!]
		\centering
		\subfloat[SiDB fabrication]{
			\includegraphics[width=.35\linewidth]{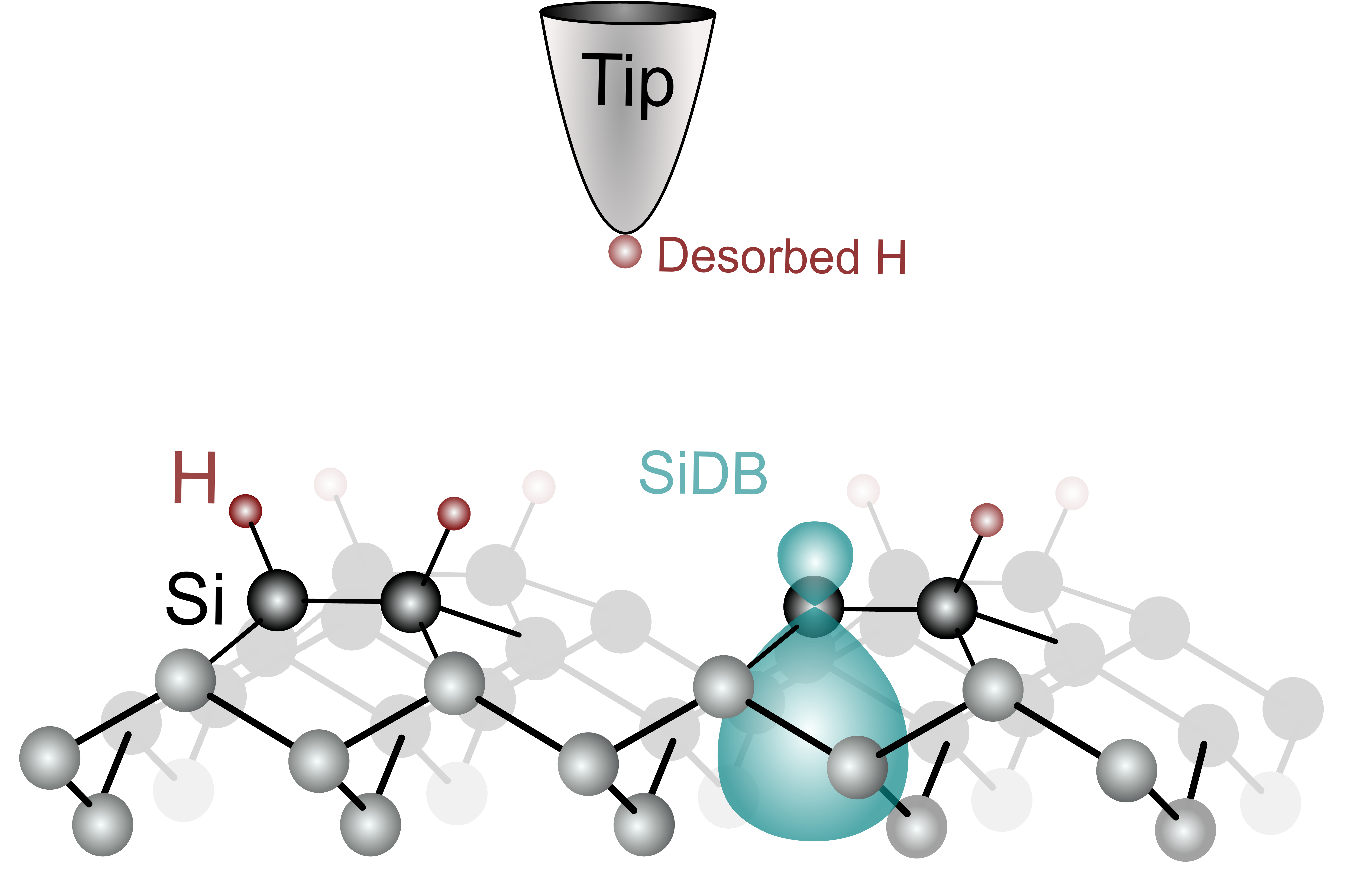}
			\label{fig:sidbs:generation}
		} \hfil
		\subfloat[Schematic lattice view]{
			\includegraphics[width=.35\linewidth]{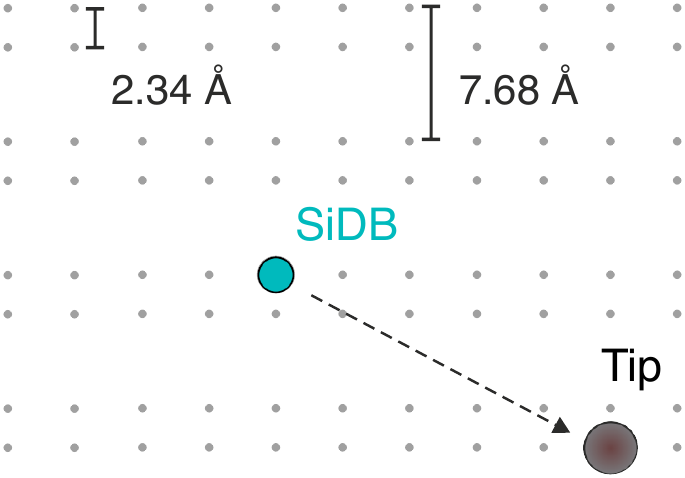}
			\label{fig:sidbs:lattice}
		} \hfil
		\subfloat[Well-separated SiDBs]{
			\includegraphics[width=.35\linewidth]{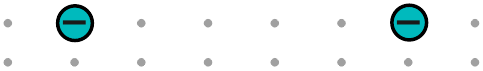}
			\label{fig:sidbs:separated}
		} \hfil
		\subfloat[Interacting SiDBs]{
			\includegraphics[width=.35\linewidth]{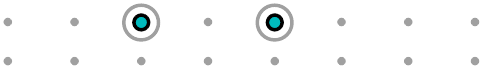}
			\label{fig:sidbs:degenerated}
		} \\
		\subfloat[BDL pair: binary \texttt{0}]{
			\includegraphics[width=.35\linewidth]{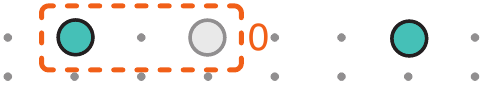}
			\label{fig:sidbs:bdl_0}
		} \hfil
		\subfloat[BDL pair: binary \texttt{1}]{
			\includegraphics[width=.35\linewidth]{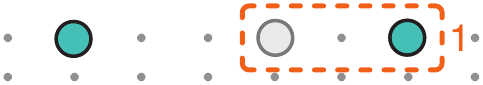}
			\label{fig:sidbs:bdl_1}
		} \\
		\subfloat[BDL wire segment transmitting a binary \texttt{1}]{
			\includegraphics[width=.8\linewidth]{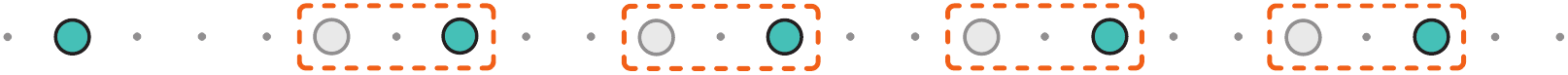}
			\label{fig:sidbs:bdl}
		} \\
		\subfloat[OR-gate (\texttt{10})]{
			\includegraphics[width=.35\linewidth]{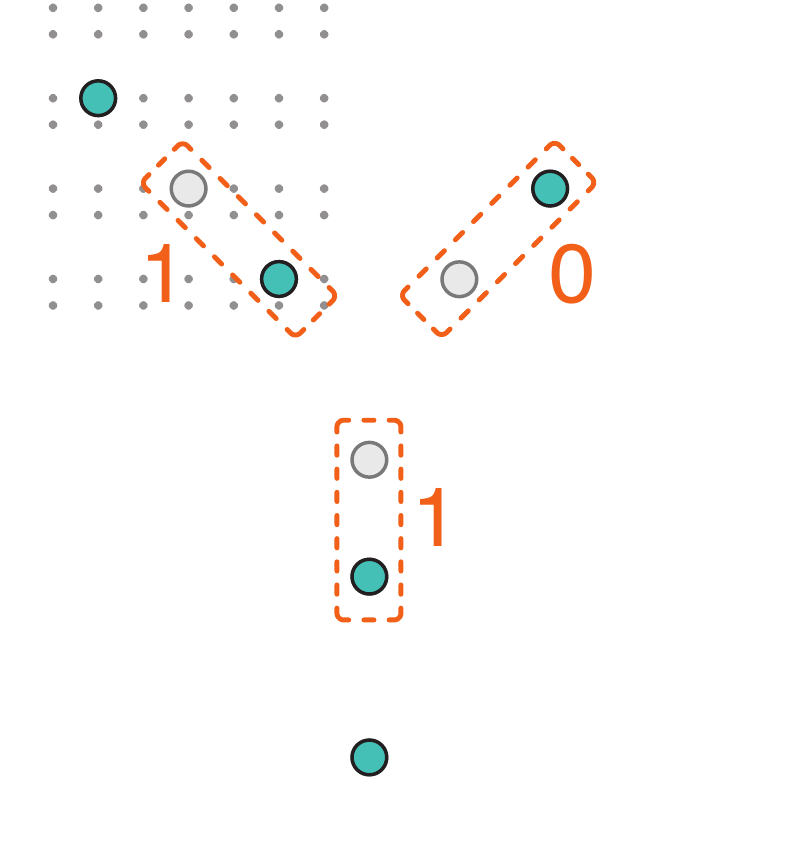}
			\label{fig:sidbs:or}
		} \hfil
		\subfloat[AND-gate (\texttt{10})]{
			\includegraphics[width=.35\linewidth]{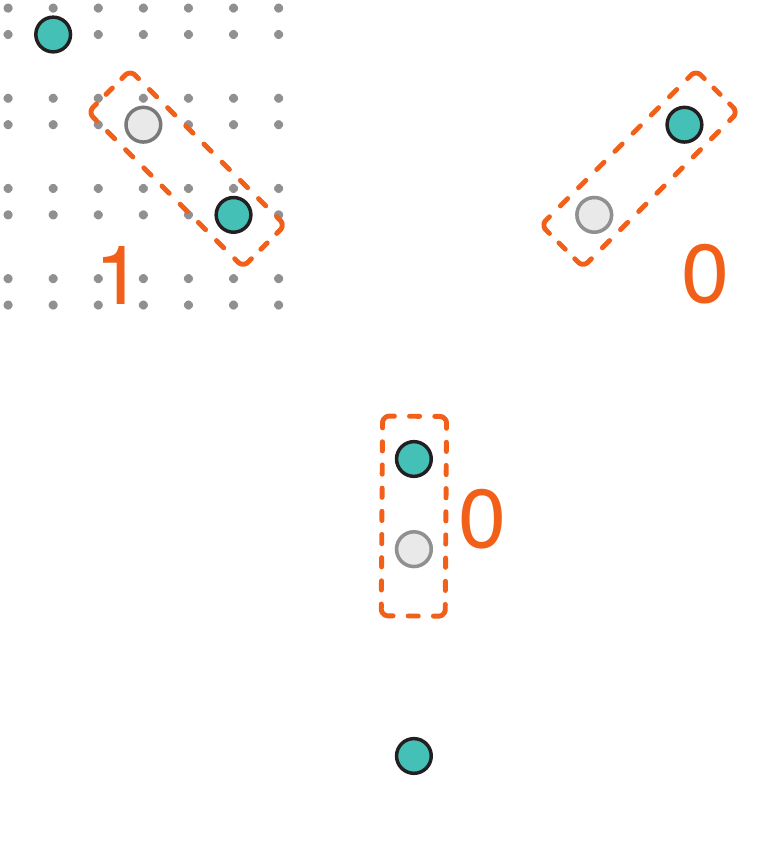}
			\label{fig:sidbs:and}
		}
		\caption{Elemental SiDB logic components}
		\label{fig:sidbs}
	\end{figure}

	
	SiDBs are fabricated on a \emph{Hydrogen-passivated Silicon}~(H-Si) surface. By using the atomically-sharp probe of a \emph{Scanning Tunneling Microscope}~(STM), a voltage can be applied that breaks the bond between silicon and hydrogen with atomic precision~\cite{pavlicek2017tip, huff2017atomic, achal2018lithography, rashidi2022automated, huff2019landscape, achal2019detecting, onoda2021ohmic, altincicek2022atomically}. The \mbox{split-off} hydrogen atom is then desorbed to the probe and leaves behind an open, \ie, dangling, valence bond; an SiDB as illustrated in \reffig{sidbs:generation}. A top view on \reffig{sidbs:generation} is  schematically shown in \reffig{sidbs:lattice}, where the teal-colored dot represents the SiDB. It also illustrates the basic principle of the fabrication process of an SiDB layout where, after the first SiDB is generated, the tip moves to a new \mbox{Si-dimer} to desorb the next hydrogen atom to generate a second SiDB. Each SiDB created this way acts as a \mbox{chemically-identical} quantum dot that can confine a maximum of two electrons due to its \mbox{$sp^{3}$-orbital}~\cite{taucer2015silicon}. Thus, each SiDB is either negatively, neutrally, or positively charged, \ie, in precisely one of three different states.
	
	Moreover, the SiDBs' charge transition energy levels lie within the bandgap: intuitively, this means that their states are stable unless disturbed by the presence of other charges~\cite{rashidi2016timeresolved, pitters2011charge}. In \mbox{n-doped} systems with a near-surface depletion region, SiDBs are able to maintain their \mbox{quantum dot} property but tend to be negatively charged due to the raised Fermi energy. This precise property is exploited in order to use SiDBs in the creation of logic elements like \mbox{gates \cite{huff2018binary}}. In fact, SiDBs can be created in pairs of \mbox{metastable} configurations such that the state of one SiDB depends on and influences the other's via band bending; a concept coined \emph{Binary-Dot Logic}~(BDL)~\cite{huff2018binary}. 
	
	\reffig{sidbs:lattice} to \reffig{sidbs:and} depict \mbox{top-down} views of SiDBs drawn as teal halos on an H-Si surface, where hydrogen atoms are visualized by gray dots. In \reffig{sidbs:separated}, two SiDBs are created with a spacing of \mbox{$\approx \SI{2}{\nm}$} such that their interaction is limited and both assume a negative charge state. In contrast, SiDBs that are placed with just a single dimer in between, as illustrated in \reffig{sidbs:degenerated}, strongly interact and, thus, share a single electron in superposition due to their symmetry if not externally excited. With the presence of an external charge, however, the electron will assume the position with maximum distance to said charge to minimize the system's total electrostatic potential energy. This concept is illustrated in \reffig{sidbs:bdl_0} and \ref{fig:sidbs:bdl_1}, where an additional SiDB is placed on the right and left, respectively, and acts as a \emph{perturber} that inflicts Coulombic pressure on the other two. This yields a BDL pair with a discrete charge state realizing binary states \texttt{0} and \texttt{1} as illustrated by the orange  \mbox{rectangles \cite{huff2018binary, ng2020siqad}}. 
	
	In a similar fashion, wires and gates can be realized. For example, a sequence of BDL pairs as depicted in \reffig{sidbs:bdl} acts as a binary wire that transmits information exclusively via the repulsion of electrical fields~\cite{huff2018binary}. The same concept can be utilized to realize, \eg, an \mbox{OR-gate} as shown in \reffig{sidbs:or} (here, illustrated using the binary input~\texttt{10}; the repulsion leads to output~\texttt{1}). Both, an SiDB wire consisting of eight SiDBs as well as an \mbox{OR-gate} (total size smaller than \SI{30}{\square\nm}) were already experimentally realized in a lab by Huff~\etal~\cite{huff2018binary}. Additionally, further gates can be realized using this concept as, \eg, the \mbox{AND-gate} shown in \reffig{sidbs:and}.

	\subsection{Physical Simulation} \label{sec:prelims:simulation}
	To explore this new emerging technology and to validate designs of novel circuits, the physical simulation of SiDB layouts is of immense interest. 		
	To predict the state that will be assumed by an arbitrary arrangement, \ie, a layout, of multiple SiDBs, an electrostatic potential simulation must be conducted. As it is known from charges in free space or dielectric materials, the electrostatic potential energy between two entities does decay with $\sfrac{1}{d}$ (with $d$ being the distance). 
	
	For SiDBs, however, an additional exponential fraction is considered on top of that. This is because electric charges 
	are shielding the electrostatic field. More precisely, the electrostatic potential $V_{i,j}$ at \mbox{position $i$} generated by an SiDB in state  $n_{j} \in \{-1,0,1\}$ at \mbox{position $j$} is given by~\cite{huff2018binary, huff2019landscape}
	\begin{align}
		V_{i,j} = -\frac{q_{e}}{4  \pi \epsilon_{0} \epsilon_{r}} \cdot \frac{e^{- \frac{{d_{i,j}}}{\lambda_\mathit{tf}}}}{d_{i,j}} \cdot n_{j},
		\label{eq:potential_sidb}
	\end{align}
	where $\lambda_{\mathit{tf}}$ defines the \emph{Thomas-Fermi screening length} and $\epsilon_{r}$ the \emph{dielectric constant} which were experimentally extracted to be \SI{5}{\nm} and $5.6$, respectively~\cite{huff2018binary}. Moreover, $\epsilon_{0}$, $q_{e}$ and $d_{i,j}$ are the \emph{vacuum permittivity}, the \emph{electron charge ($q_{e}=-e$; $e$: elementary charge)}, and the \emph{Euclidean distance} between position $i$ and $j$, respectively. As can be seen from \refequa{potential_sidb}, all SiDB states influence each other and, by this, form an interwoven system. A layout's total electrostatic potential \mbox{energy $E$} is then calculated by using the superposition principle of the electrostatic potential~\cite{griffiths_2017}, \ie , 
	\begin{align}
		E =  -\sum_{i<j} V_{i,j} \cdot n_{i} \cdot q_{e}.
		\label{eq:energy}
	\end{align}
	The expected state of an SiDB layout is given by the metastable charge configuration with the lowest energy. Therefore, in order to determine the said state, \refequa{energy} has to be minimized---yielding a \mbox{high-dimensional} optimization problem.
	
	On top of that and in order to guarantee a valid simulation result, the physical constraint of \emph{metastability} must be met. Metastability can be described by the combination of two criteria, namely \emph{population stability} and \emph{configuration stability}. Both must be obeyed to guarantee simulation result validity.
	
	\paragraph{Population Stability} 
	The charge state of each SiDB must be consistent with the energetic charge transition levels $\mu_{-}$ and $\mu_{+}$ relative to the Fermi energy. The energy of the SiDB's charge transition level at position~$i$ depends on the local electrostatic potential $V_{\mathit{local},i}$ at that position defined by
	\begin{align}
		V_{local,i} = \sum_{j, j \neq i} V_{i,j}. 
		\label{eq:local_potential}
	\end{align}
	Intuitively, this means that an SiDB is preferably neutrally charged when adjacent SiDBs are negatively charged and vice versa. This relation is formally expressed by the following conditions: SiDB- when $\mu_{-} + V_{\mathit{local},i} \cdot q_{e} < 0$, SiDB+ when $ \mu_{+} + V_{\mathit{local},i} \cdot q_{e} > 0 $, and SiDB0 otherwise.
	
	\paragraph{Configuration Stability}
	If there is no feasible \mbox{single-electron} hop event between arbitrary SiDBs leading to a lowering of the system's total electrostatic potential energy, the charge state fulfills the configuration stability. In other words, if this constraint is not satisfied, there would exist a state of lower electrostatic potential energy that the system would spontaneously converge to.
	
	\smallskip
	
	Overall, the charge configuration with the lowest electrostatic potential system energy that is also physically valid, \ie, satisfies metastability, is called the system's \emph{ground state}. It represents the charge configuration of a given SiDB system at low temperature and, by this, its physical behavior. Hence, the goal of any physical simulation technique is to determine the said ground state for a given SiDB layout. 
	
	
	\section{Motivation} \label{sec:simulations:approaches}
	In this section, SiDB simulation approaches from the literature are reviewed first. Afterward, the obstacles that arise when attempting to balance the runtime and accuracy of these algorithms are discussed---providing the motivation for this work. 
	
	\subsection{Previously Proposed Simulation Approaches}
	\label{sec:prelims:algorithms}
	
	As outlined above, the physical simulation of SiDB logic poses a \mbox{high-dimensional} optimization problem that, due to its exponential complexity, is computationally intractable. Since simulation constitutes a core capability of any design automation flow for \mbox{SiDB-based} logic, managing this complexity is a key priority. However, only a few automatic solutions for the physical simulation of SiDBs exist so far, and they are either of high runtime/poor scalability (due to the exponential search space) or of a rather poor accuracy (due to aggressive approximations). Two methods represent the current state of the art:
	\emph{ExhaustiveGS}~(ExGS)~\cite{vieira2022three, ng2020thes} and \emph{SimAnneal}~\cite{ng2020siqad}. 
		
		\subsubsection{ExGS}
		Since every SiDB can either be negatively, neutrally, or positively charged, a layout consisting of \mbox{$n$ SiDBs} can exhibit~$3^n$ unique charge configurations. ExGS enumerates all of them, determines if they are physically valid, and, if this is the case, computes the electrostatic potential energy of the layout via \refequa{energy}. This eventually allows obtaining the charge configuration with minimal energy.
		
		Since ExGS covers the entire search space exhaustively, it is guaranteed to always determine the system's ground state. At the same time, however, this requires the traversal of the entire exponential search space and, hence, even in the best case takes an exponential runtime to do so. Thus, ExGS is practically applicable to small layouts only.
		
		\subsubsection{SimAnneal} 
		In contrast to ExGS, SimAnneal is a heuristic algorithm that generates approximate results. As the name suggests, it employs the strategy of \emph{simulated annealing}~\cite{kirkpatrick1983optimization}, which is a probabilistic technique for approximating the global optimum of a given objective function 
		which, in this case, is the electrostatic potential energy of closely-placed SiDBs. More precisely, the algorithm consists of two components, namely \emph{surface hopping} and \emph{population control}.

		In the surface hopping phase, a fixed number of electrons and holes can hop to neutral SiDBs. To this end, each hop is subject to an \mbox{Arrhenius-like} acceptance function with $\Delta E$ being the change in \refequa{energy} resulting from the hop. 
		At the beginning of the annealing process, the artificial thermal energy $k_{B}T$ starts at a high value and decreases with the number of annealing cycles. Hence, over time, \mbox{energetically-high} charge configurations are becoming increasingly unlikely. As a consequence, the system eventually locks into a final charge configuration.
		
		However, since the total number of electrons in the system is not fixed, but, instead, to be determined by the algorithm in question, surface hopping is not sufficient to determine the ground state. Hence, the population control phase is additionally needed. Here, electrons move between SiDBs and an imaginary electron reservoir, \ie, the silicon bulk. The probability of an electron moving from the reservoir to the SiDB or vice versa is described by the \mbox{\emph{Fermi-Dirac statistics} \cite{peter2010fundamentals}}. At the end of the annealing schedule, all simulation results are evaluated regarding the population and configuration stability and the energetically lowest one is returned.
		
		SimAnneal is a powerful algorithm that is capable of correctly determining the SiDBs' charge distribution of many layouts. Moreover, since SimAnneal is a \mbox{physically-inspired} algorithm, it inherently captures the charge configuration of SiDBs. On the downside, it is a probabilistic heuristic that requires many simulation steps to result in the ground state. Especially, when many charge configurations lead to a similar electrostatic potential system energy close to the ground state, long runtimes are required. Hence, SimAnneal's biggest drawback is that it does not prune its search space. 
		Even though the number of potential states shrinks during its runtime in accordance with the artificial thermal energy, the chances of getting stuck in a local minimum directly correlate with input size. 
		Consequently, high initial temperatures are necessitated which negatively impacts the runtime.
		
		\subsection{Resulting Problem: Trade-off between Efficiency and Accuracy} \label{sec:trade_off}
		
		As elaborated on above, the goal of physical simulation algorithms for SiDB layouts is to determine the system's ground state. Probabilistic heuristics like SimAnneal are usually run a number of times on the same input hoping that the ground state is determined at least once. To evaluate their performance in doing so, their required runtime could be used as a measure. However, in probabilistic algorithms, \emph{single runtime}~$\mathit{t_{s}}$ and \emph{result accuracy}~$\mathit{p_{s}(t_{s})}$ are strongly correlated, \ie, more single runs of, \eg, SimAnneal, increase the likelihood of determining a layout's ground state and vice versa. Hence, single runtime alone is not a meaningful metric as accuracy should also be considered. This trade-off is captured by the \emph{Time-To-Solution}~(TTS) metric that provides the \emph{total} runtime needed to derive a solution of a certain confidence level~$\eta$ \cite{albash2018demonstration}. More precisely, TTS is defined by
		\begin{align}
			\mathit{TTS(t_{s})} &= \mathit{t_{s}} \cdot \mathit{R(t_{s})}, \text{with} \\
			\mathit{R(t_{s})} &= \begin{cases}
				\frac{\ln(1-\eta)}{\ln(1-p_{s}(t_{s}))} & p_{s}(t_{s}) \in (0,\eta) \\
				1 & p_{s}(t_{s}) \in [\eta,1] 
			\end{cases}.
		\end{align}

		ExGS constitutes the extreme case of this \mbox{trade-off:} The single runtime is maximal (as the \emph{entire} search space is traversed), but the obtained accuracy is 100\,\% (since all possibilities are considered). Thus, ExGS' TTS is large compared to SimAnneal's. In contrast, SimAnneal's accuracy is lower than ExGS', whereas its single runtime is smaller by several orders of magnitude. Since the single runtime influences the accuracy, small runtimes lead to a small accuracy and vice versa. Hence, an ideal \mbox{trade-off} between these two attributes is desired to yield an efficient simulation. Overall, neither of the existing solutions---ExGS or SimAnneal---provide such a satisfactory \mbox{trade-off} between efficiency and accuracy. In this work, we address this shortcoming.

		
		

		\section{Proposed Algorithm - \emph{QuickSim}} \label{sec:proposed}
		In this section, we propose a novel algorithm 
		for physical simulation of SiDB layouts. We start by sketching the general idea on a more abstract level. Afterward, we go into more detail 
		and present the algorithm implementing the proposed idea. 
		
		\subsection{General Idea} \label{sec:prelims:idea}
		
		
		
		The proposed algorithm \emph{QuickSim} for physical SiDB simulation aims at providing a significantly improved \mbox{trade-off} between single runtime and result accuracy. To this end, 
		we are making use of effective search space pruning by incorporating ideas from statistical methods like \mbox{\emph{max-min diversity distributions} \cite{resende2010grasp}}. This leads to a new simulation algorithm that works fundamentally differently than the state of the art. 
		
		More precisely, while previously proposed approaches alter the given system's electron number several times during one simulation run, considering \emph{all} possible charge configurations in the case of ExGS or effectively randomly guessing configurations in the case of SimAnneal, we propose to iteratively and purposefully distribute electrons on the basis of insights provided by physical principles. 
		Given the electrons' tendency to always occupy SiDBs that are placed the farthest distance away from one another (to minimize their electrostatic potential energy), we can infer that any given number of electrons must settle in an \mbox{equally-spaced} allocation across the SiDBs. In other words, the electrostatic potential energy of a system is \emph{minimized} when the electron distribution's diversity is \emph{maximized}.
		
		We can create such \mbox{max-min} diversity distributions by iteratively filling up placed SiDBs with electrons starting from the neutral state such that each newly introduced charge has a maximal distance from all others. Conducting this iteration once with every SiDB as the starting point and, along the way, computing both the system's electrostatic potential energy and metastability, we obtain an algorithm that is less probabilistic and less sensitive to hyperparameters than SimAnneal. Additionally, it requires fewer simulation steps to achieve the same accuracy, which considerably reduces its TTS. 
		
\begin{figure}[t!]
			\centering
			\subfloat[Occupying initial SiDB]{
				\includegraphics[width=.35\linewidth]{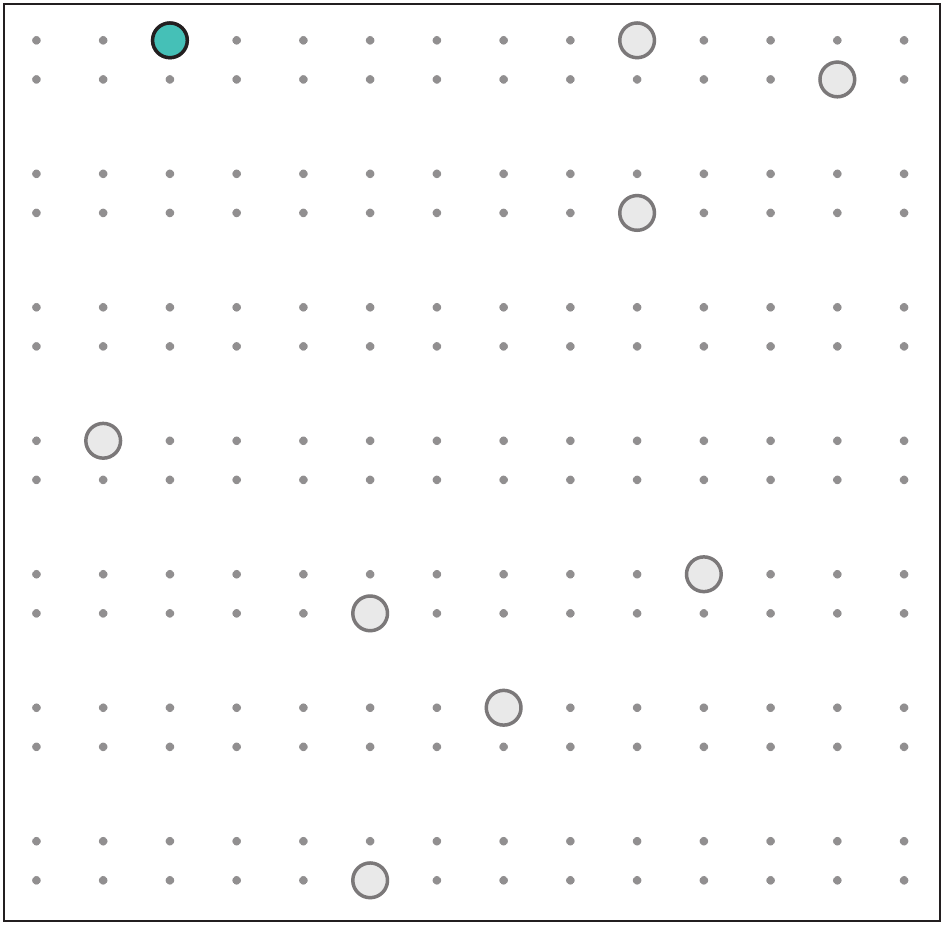}
				\label{fig:initial}
			} \hfil
			\subfloat[Occupying \nth{2} SiDB]{
				\includegraphics[width=.35\linewidth]{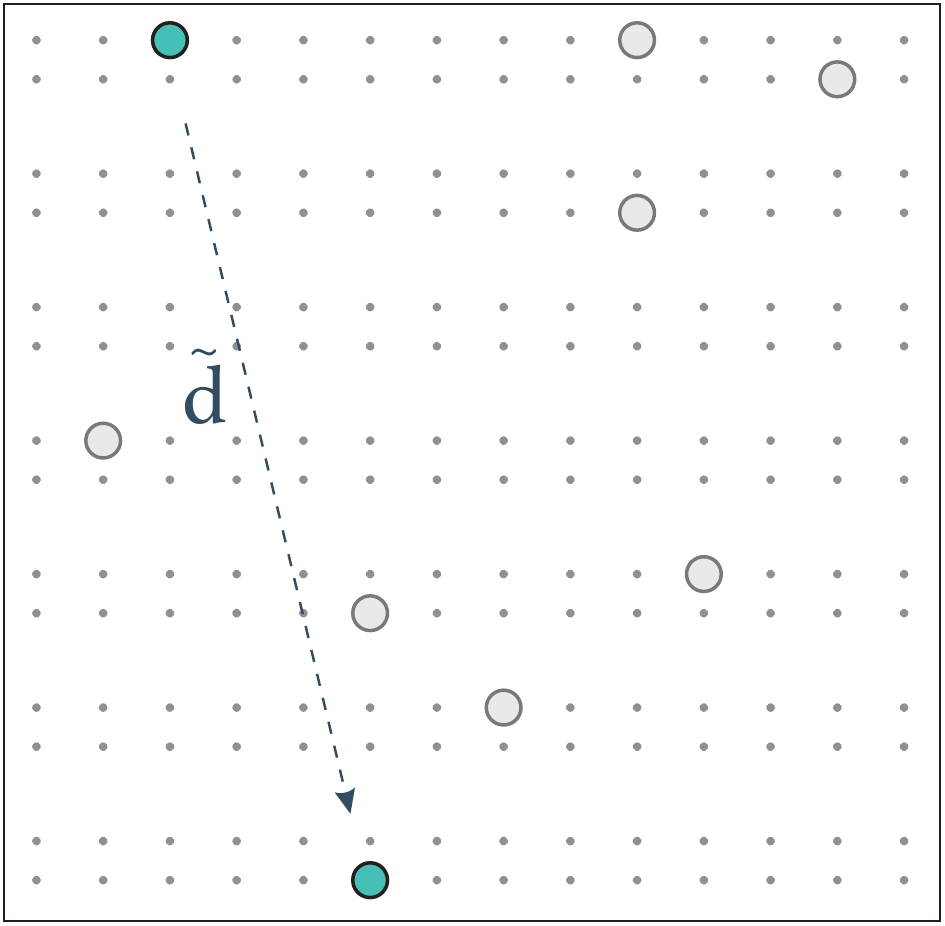}
				\label{fig:first_iteration}
			} \\
			\subfloat[Occupying \nth{3} SiDB]{
				\includegraphics[width=.35\linewidth]{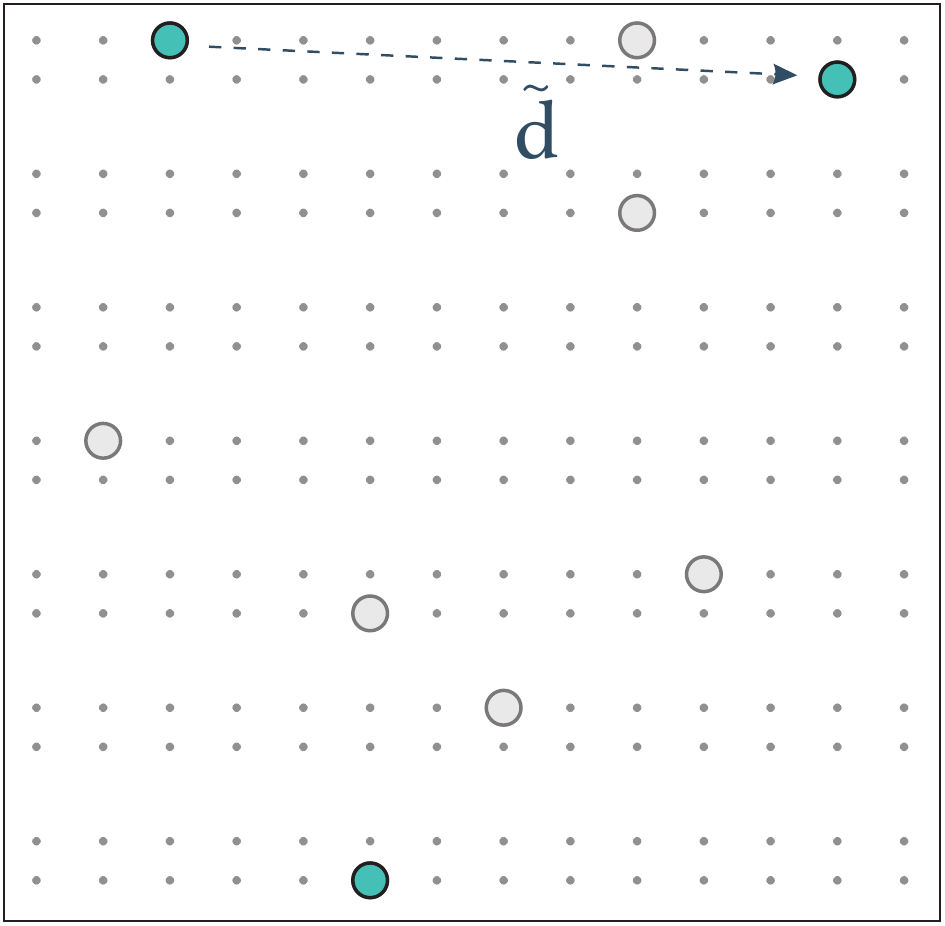}
				\label{fig:second_iteration}
			} \hfil
			\subfloat[System is metastable]{
				\includegraphics[width=.35\linewidth]{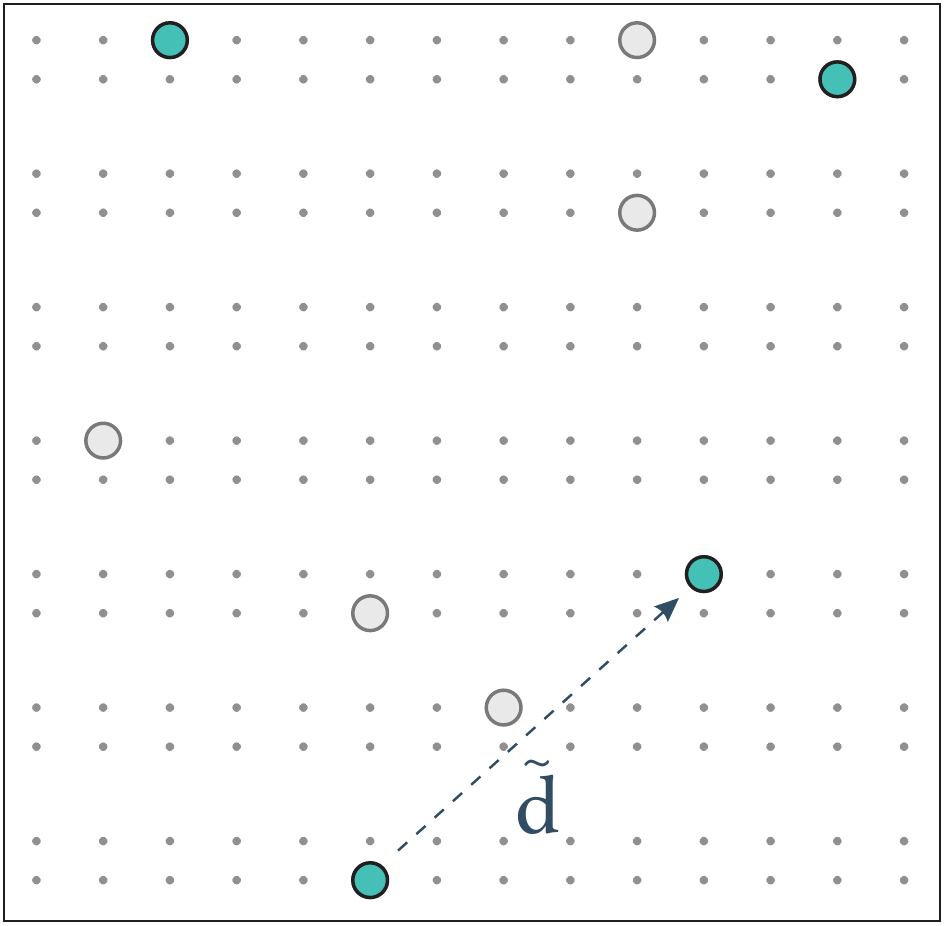}
				\label{fig:metastable}
			}
			\caption{Illustration of the general idea ($\mu_{-} = \SI{-0.15}{\eV}$)}
			\label{fig:illustration_algorithm}
		\end{figure}

		For example, in order to simulate the layout shown in \reffig{initial} (consisting of nine SiDBs), a total of $3^9 = 19\,683$ charge configurations have to be considered in the worst case (as done by ExGS). SimAnneal allows reducing this number, but, due to the intrinsic probabilistic characteristic of this approach, still, several hundreds of charge configurations are considered.
		For each charge configuration, the system's electrostatic potential energy and the metastability condition have to be determined and checked, respectively---yielding substantial computational costs. In contrast, only focusing on charge configurations including the most distant SiDBs---as proposed in this work and illustrated by \reffig{initial} to \reffig{metastable}---allows to reduce this number to $9 \cdot 4 = 36$ total charge configurations only. That is, rather than hundreds of charge configurations, just a fraction of them are considered while, at the same time, almost no optimal solution is pruned---evidently a great \mbox{trade-off} between efficiency and accuracy.
		
		
		\subsection{Implementation Details}

		\begin{algorithm}[!t]
			\SetAlgoLined
			\DontPrintSemicolon
			\KwIn{SiDB Layout $L$ comprised of $n$ dangling bonds}
			\KwIn{Physical parameters $P = \{ \mu_{-}, \mu_{+}, \lambda_{\mathit{tf}}, \epsilon_{r} \}$}
			\KwIn{Number of cycles $c$}
			\KwOut{Electron distribution $D$ and system energy $E$}
			$E^* \gets \infty$\;
			$D^* \gets [D_0 = 0, \dots, D_n = 0]$ \label{line:max-distance:empty-D}\;
			\For{$\mathit{cycle} \gets 1$ \KwTo $c$\label{line:max-distance:cycles}}
			{
				\ForEach{$\mathit{db} \in L$\label{line:max-distance:foreach_sidb}}
				{
					$D \gets [\mathit{db} = -1]$\label{line:max-distance:starting_sidb} \;
					\For{$i \gets 1$ \KwTo $n - 1$\label{line:max-distance:iteration_increased}} 
					{
						$D \gets \textsc{Adjacent-Search}(L, D)$\label{line:max-distance:adjacent-search} \;
						$E \gets \text{energy of } L \text{ given } D \text{ and } P$\tcp*{\refequa{energy}} \label{line:max-distance:energy}
						\If{$D \text{ is phys. valid given } P$ \textbf{\emph{and}} $E < E^*$\label{line:max-distance:valid}}
						{
							$D^* \gets D$ \label{line:max-distance:best-d}\;
							$E^* \gets E$ \label{line:max-distance:best-e}\;
							\textbf{break}\;
						}
					}
				}
			}
			\Return $(D^*, E^*)$ \label{line:max-distance:return} \; 
			\caption{Physical Simulation}
			\label{algo:max-distance}
		\end{algorithm}

		\begin{algorithm}[!t]
			\SetAlgoLined
			\DontPrintSemicolon
			\KwIn{SiDB Layout $L$ comprised of $n$ dangling bonds}
			\KwIn{Partial electron distribution $D$}
			\KwOut{Extended electron distribution $D'$}
			$d^* \gets \{ \min_{p \in D}{\mathit{dist}(u, p)} \ | \ u \in L \setminus D \}$ \label{line:adjacent_search:min} \;
			$\tilde{d} \gets \max_{\mathit{d} \in d^*} d$\; \label{line:adjacent_search:max}
			$S \gets \{u \ | \ u \in L \setminus D, d^*_{u} \geq \tilde{d} \cdot \alpha \}$ \label{line:adjacent_search:alpha} \; 
			Randomly select one $\mathit{db}' \in S$\; \label{line:adjacent-search:select_randomly}
			\Return $D \cup [\mathit{db}' = -1$] \label{line:adjacent-search:return} \;
			\caption{Adjacent-Search\label{algo:adjacent-search}}
			\label{algo:adjacent_search}
		\end{algorithm}
		

		
		
		The details of \emph{QuickSim} are exemplified given the pseudocode from \refalgo{max-distance} and~\ref{algo:adjacent-search} and visualized again in \reffig{illustration_algorithm}. 
		The algorithm starts with an empty electron distribution~$D$, \ie, all SiDBs in the given layout $L$ are set to be neutrally charged (\refalgo{max-distance}, \refline{max-distance:empty-D}). For each SiDB in the layout as the starting point (\refalgo{max-distance}, \refline{max-distance:foreach_sidb}), electrons are successively distributed while maximizing diversity. To this end, the initial SiDB $\mathit{db}$ is assigned an electron, \ie, flipped to a negative charge (\refalgo{max-distance}, \refline{max-distance:starting_sidb}). From there, \refalgo{adjacent_search} is iteratively called (\refalgo{max-distance}, \refline{max-distance:adjacent-search}) to determine other SiDBs with maximum distance to all negatively charged ones (\refalgo{adjacent_search}, \reflines{adjacent_search:min}{adjacent_search:max}). These are enumerated within a threshold given by the value $\alpha$ (\refalgo{adjacent_search}, \refline{adjacent_search:alpha}). While $\alpha$ is represented as a variable factor between $0$ and $1$, the algorithm was found to perform best when picking $\alpha = 0.7$. Out of the possible candidates, one SiDB is randomly selected (\refalgo{adjacent_search}, \refline{adjacent-search:select_randomly}) and assigned an electron (\refalgo{adjacent_search}, \refline{adjacent-search:return}). The newly resulting electron distribution $D$ in the layout~$L$ is utilized to compute its electrostatic potential energy $E$ under the physical parameters $P$, \ie, $\mu_{-}$, $\mu_{+}$, $\lambda_{\mathit{tf}}$ and $\epsilon_{r}$ (\refalgo{max-distance}, \refline{max-distance:energy}). If $D$ is physically valid under $P$ (cf.~\refsec{prelims:simulation}) and if $E$ is smaller than the best energy value found so far (\refalgo{max-distance}, \refline{max-distance:valid}), this constitutes a better candidate for the overall simulation result, thus, both~$D$ and~$E$ are kept as current best (\refalgo{max-distance}, \reflines{max-distance:best-d}{max-distance:best-e}).
		
		This entire procedure is repeated $c$ times (\refalgo{max-distance}, \refline{max-distance:cycles}) and the best result found is returned (\refalgo{max-distance}, \refline{max-distance:return}).
		
		\section{Experimental Evaluations} \label{sec:eval}
		
		To demonstrate the applicability and performance of \emph{QuickSim} in comparison to the state of the art, we implemented \emph{QuickSim} and conducted exhaustive experimental evaluations. This section summarizes and analyzes our findings. To this end, we first describe the experimental setup. Afterward, we provide the respectively obtained results from experimental simulations considering three different classes of SiDB layouts, namely \begin{enumerate*}
			\item randomly-generated,
			\item \mbox{machine learning (ML)-generated}, and
			\item established ones from the literature.
		\end{enumerate*}
		Finally, the 
		obtained results are discussed. 
		
		%

		\subsection{Experimental Setup} \label{sec:eval:setup}
		
		In order to properly evaluate the \mbox{trade-off} between efficiency and accuracy as well as to showcase how \emph{QuickSim} overcomes this, we performed \emph{Time-To-Solution}~(TTS, as defined in \refsec{trade_off}) evaluations. To this end, we considered  the two \mbox{state-of-the-art} physical simulation algorithms (ExGS and SimAnneal as reviewed in \refsec{prelims:algorithms}) as well as \emph{QuickSim}. As benchmarks, we considered a variety of SiDB layout classes whose details are reviewed in the following section. To determine $p_{s}(t_{s})$ as introduced in \refsec{trade_off}, the simulations have been conducted $10\,000$ times and the success number (of correct simulation results) was collected. Thus, \mbox{$p_{s}(t_{s}) = \sfrac{\mathit{\#success}}{10\,000}$}. For the whole evaluation, the confidence level has been set to $\eta = 99.7\,\%$.
		
		For comparisons, the SimAnneal implementation as provided by \siqad~\cite{ng2020siqad} was utilized, whereas ExGS and \emph{QuickSim} are custom implementations. All code has been implemented on top of the \href{https://github.com/marcelwa/fiction}{\emph{fiction}} toolset and compiled with AppleClang 14.0.0. All obtained results are publicly available at \href{https://github.com/cda-tum/sidb-layouts-quicksim-validation}{GitHub}. All experiments were performed on a macOS $12.6$ machine with an Apple Silicon M1 Pro SoC with 32\,GB of integrated main memory. 
		
		Since SimAnneal offers a range of hyperparameters that can be tuned to a problem instance at hand, we considered the TTS data for SimAnneal in two configurations: \emph{default}, \ie, as set out-of-the-box, and \emph{optimized}, \ie, manually tuned settings for improved TTS on the considered benchmark set. Thus, \emph{QuickSim's} performance can be compared against a setting that provides an even better simulation performance than supplied by~\cite{ng2020siqad}. It turns out that the variation in the number of annealing cycles strongly correlates with the TTS. In contrast, the other hyperparameters do not show any considerable influence on the TTS. Hence, the TTS is optimized by tuning the number of annealing cycles. For the benchmark layouts, manual tuning was conducted to achieve the smallest TTS values. It can be concluded that reducing the number of annealing cycles by one order of magnitude from $10\,000$ (def.) to $1\,000$ (opt.), results in optimized TTS values---providing a highly optimized baseline against which we compare \emph{QuickSim} in this work. For both ExGS and \emph{QuickSim} we apply default parameters, for ExGS that entails using \mbox{two-state} simulation\footnote{Three-state simulation is not required because no layout in the benchmark set has an \mbox{inter-dot} spacing of $< \SI{7}{\nano\meter}$ and, thus, cannot exhibit positive charge states.} and for \emph{QuickSim} $\alpha = 0.7$ and $c = 80$.

		\subsection{Considered Layouts and Obtained Results} \label{sec:eval:random}
		
		
		In the first series of evaluations, we considered \mbox{randomly-generated} gate layouts. While the simulation of readily available gate libraries has more practical value (and, hence, is considered as well below), this allows demonstrating the performance of the respective simulations for potential future gate libraries showcasing the broad applicability. To this end, we considered the \mbox{2-in-1-out} and \mbox{2-in-2-out-templates} that have recently been proposed in~\cite{walter2022hexagons} and are considered promising SiDB candidates for fabrication. Our benchmark set contains 114 random implementations for each template with a range of SiDBs per layout from 20 to 25.\footnote{As mentioned above, SiDB spacing of less than \SI{7}{\nano\meter} is prohibited to avoid degenerate states and over-saturated potential landscapes.} 

	In a second series of evaluations, we considered SiDB layouts generated using the machine learning approach presented in~\cite{lupoiu2022automated}. In contrast to the \mbox{randomly-generated} instances, these implement common logic operations such as AND, OR, NAND, NOR, etc. Hence, these instances provide a representative use case for physical simulation applied in the evaluation of new gate libraries. 
	This benchmark set contains a total of $148$ novel SiDB layouts. As above, we simulated them with all three approaches and compared their TTS. To this end, each gate has been simulated in up to five configurations: with input perturbers specifying one of the possible~$2^n$ input combinations as well as without any input perturbers at all. However, some ML-generated gates are very sensitive to a change in physical parameters and sometimes do not work correctly for all input combinations (\texttt{00,01,10,11}). Hence, a further series of evaluations are conducted. 
	
	In a third series of evaluations, all $12$ gate layouts from~\cite{walter2022hexagons} are simulated. They represent a library of established gates in the SiDB domain. The simulation is conducted identically to the one in the evaluation of the \mbox{ML-generated} layouts. 
	
	All obtained results are summarized in \reftab{random_gates}, \reftab{ML_gates}, and \reftab{established_gates} for the randomly-generated, \mbox{ML-generated}, and established layouts, respectively. In each table, the first three columns provide the name of the benchmark, their SiDB range, and the number of simulated layouts (instance count). Afterward, the accuracy and the accumulated TTS values (in sec.) for all considered simulation approaches (and their configurations) are stated. Since ExGS is an exact algorithm and, thus, always returns the ground state, the accuracy is not listed in the table to avoid redundancy. Furthermore, to ensure a sophisticated evaluation of \emph{QuickSim}, TTS gained with optimized settings is collected in addition to the TTS of SimAnneal in default settings as discussed before providing a highly optimized baseline against which we compare \emph{QuickSim}. 
	In the tables' last two rows, the mean value of the accuracies and the sums of the individual accumulated TTS are collected, respectively.

	\begin{table}[!t]
		\centering
		\caption{Physical simulation of randomly-generated layouts}
		\label{tab:random_gates}
		\begin{minipage}{0.95\linewidth}
			\centering
			\begin{adjustbox}{max width=\linewidth}
				\begin{tabular}{cccccccrccrccr}
					\toprule
					\multicolumn{3}{c}{\multirow{2}{*}{\textsc{Benchmark}}} &  \phantom{m} & \multicolumn{10}{c}{\textsc{Physical Simulation}} \\
					\cmidrule{5-14}
					\multicolumn{4}{c}{~} &  \multicolumn{1}{c}{\textsc{ExGS~\cite{ng2020thes}}} & & 
					\multicolumn{5}{c}{\textsc{SimAnneal~\cite{ng2020siqad}}} & & \multicolumn{2}{c}{\textsc{\emph{QuickSim}}}  \\
					\cmidrule{1-3} \cmidrule{5-5} \cmidrule{7-11} \cmidrule{13-14}				\multicolumn{1}{c}{\multirow{2}{*}{Name}} & \multicolumn{1}{c}{\multirow{2}{*}{\#SiDBs}} & \multicolumn{1}{c}{\multirow{2}{*}{\#inst.}}  &  &  \multicolumn{1}{c}{\multirow{2}{*}{TTS}} & & \multicolumn{2}{c}{\multirow{1}{*}{default}} & &  \multicolumn{2}{c}{\multirow{1}{*}{optimized}} && \multicolumn{2}{c}{\multirow{1}{*}{default}}  \\
					\cmidrule{7-8} \cmidrule{10-11} \cmidrule{13-14}
					\multicolumn{6}{c}{~} &  \multicolumn{1}{c}{acc.} & \multicolumn{1}{c}{TTS} &&   \multicolumn{1}{c}{acc.} &  \multicolumn{1}{c}{TTS} &&  \multicolumn{1}{c}{acc.} & \multicolumn{1}{c}{TTS} \\
					\midrule
					\multirow{1}{*}{2-1}   
					& $20-23$ & 114 && $172.31$ & & $40.6$ & $29.59$ && $25.8$ & $11.43$ && $82.6$ & $1.18$
					\\ \addlinespace[0.5em]
					\multirow{1}{*}{2-2}   
					& $21-25 $ & 114 && $452.09$ & &  $16.1$ & $363.52$  && $11.9$ &$28.72$&& $79.7$ & $1.73$ \\
					\midrule 
					\addlinespace[0.5em]
					\multirow{1}{*}{\emph{Mean}}
					&&&& &&$28.4$ &&& $18.9$  & &&$\mathbf{81.2}$ &  \\ 
					\addlinespace[0.5em]
					\multirow{1}{*}{\emph{Total}}
					&&&&$624.40$ &&&$393.11$ &&& $40.15$&&&$\mathbf{2.91}$  \\ 
					
					
					\bottomrule
				\end{tabular}
			\end{adjustbox}
		\end{minipage}
	\end{table}

	\begin{table}[!t]
		\centering
		\caption{Physical simulation of ML-generated layouts}
		\label{tab:ML_gates}
		\begin{minipage}{0.95\linewidth}
			\centering
			\begin{adjustbox}{max width=\linewidth}
				\begin{tabular}{lcccrcrrcrrcrr}
					\toprule
					\multicolumn{3}{c}{\multirow{2}{*}{\textsc{Benchmark}}} &  \phantom{m} & \multicolumn{10}{c}{\textsc{Physical Simulation}}  \\
					\cmidrule{5-14}
					\multicolumn{4}{c}{~} &  \multicolumn{1}{c}{\textsc{ExGS~\cite{ng2020thes}}} & & 
					\multicolumn{5}{c}{\textsc{SimAnneal~\cite{ng2020siqad}}} & & \multicolumn{2}{c}{\textsc{\emph{QuickSim}}}  \\
					\cmidrule{1-3} \cmidrule{5-5} \cmidrule{7-11} \cmidrule{13-14}				\multicolumn{1}{c}{\multirow{2}{*}{Name}} & \multicolumn{1}{c}{\multirow{2}{*}{\#SiDBs}} & \multicolumn{1}{c}{\multirow{2}{*}{\#inst.}}  &  &  \multicolumn{1}{c}{\multirow{2}{*}{TTS}} & & \multicolumn{2}{c}{\multirow{1}{*}{default}} & &  \multicolumn{2}{c}{\multirow{1}{*}{optimized}} && \multicolumn{2}{c}{\multirow{1}{*}{default}}  \\
					\cmidrule{7-8} \cmidrule{10-11} \cmidrule{13-14}
					\multicolumn{6}{c}{~} &  \multicolumn{1}{c}{acc.} & \multicolumn{1}{c}{TTS} &&   \multicolumn{1}{c}{acc.} &  \multicolumn{1}{c}{TTS} &&  \multicolumn{1}{c}{acc.} & \multicolumn{1}{c}{TTS} \\
					\midrule
					\multirow{1}{*}{OR}   
					& $ 19-23 $ & 68 & & $2680.08$ &&  $59.8$ & $12.52$ &&$38.9$& $10.23$ &&$86.0$& $0.54$
					\\ \addlinespace[0.5em]
					\multirow{1}{*}{AND}   
					& $ 25 $ & 25 & & $2086.82$ & & $18.1$ & $27.59$ &&$15.0$& $21.47$ &&$66.4$& $1.60$
					\\ \addlinespace[0.5em]
					\multirow{1}{*}{NOR}         
					& $20-23 $ & 20 & & $1312.83$ && $26.0$ & $10.55$ &&$15.4$& $2.72$ &&$99.6$& $0.05$
					\\ \addlinespace[0.5em]
					\multirow{1}{*}{NAND}   
					& $20-23 $ & 16 & & $1300.58$ && $45.7$ & $7.79$ &&$37.3$& $22.12$ &&$94.9$& $0.07$
					\\ \addlinespace[0.5em]
					\multirow{1}{*}{CX}   
					& $ 29 $ & 4 & & $577.00$ && $21.5$ & $21.45$ &&$13.2$& $12.52$ &&$29.0$& $0.56$
					\\ \addlinespace[0.5em]
					\multirow{1}{*}{HA}   
					& $ 31 $ & 4& & $433.91$ &&  $17.8$ & $20.03$ &&$15.1$& $3.98$ &&$17.5$& $4.45$
					\\ \addlinespace[0.5em]
					\multirow{1}{*}{FO2}   
					& $ 27 $ & 2& & $65.42$ &&  $21.4$ & $0.78$ &&$22.8$& $0.12$ &&$23.3$& $0.38$  \\
					
					\addlinespace[0.5em]
					\multirow{1}{*}{XOR}          
					& $21-23 $ & 5 & & $4.49$ && $30.0$ & $2.09$ &&$32.4$& $0.28$ &&$95.0$& $0.02$
					\\ \addlinespace[0.5em]
					\multirow{1}{*}{XNOR}   
					& $25 $ &4& & $2.97$ && $20.4$ & $6.54$ &&$25.0$& $8.94$ &&$72.8$& $0.16$
					\\ 
					\midrule
					\multirow{1}{*}{\emph{Mean}}  &  & &  & & & $29.0$ &&& $23.9$  &&& $\mathbf{65.0}$ \\
					\addlinespace[0.5em]
					\multirow{1}{*}{\emph{Total}}
					& & & & $8464.07$ && & $109.34$  &&& $82.38$ &&& $\mathbf{7.83}$  \\
					
					\bottomrule
				\end{tabular}
			\end{adjustbox}
		\end{minipage}
	\end{table}

	\begin{table}[!t]
		\centering
		\caption{Physical simulation of established gate layouts}
		\label{tab:established_gates}
		\begin{minipage}{0.95\linewidth}
			\centering
			\begin{adjustbox}{max width=\linewidth}
				\begin{tabular}{lcccrcrrcrrcrr}
					\toprule
					\multicolumn{3}{c}{\multirow{2}{*}{\textsc{Benchmark}~\cite{walter2022hexagons}}} &  \phantom{m} & \multicolumn{10}{c}{\textsc{Physical Simulation}} \\
					\cmidrule{5-14}
					\multicolumn{4}{c}{~} &  \multicolumn{1}{c}{\textsc{ExGS~\cite{ng2020thes}}} & & 
					\multicolumn{5}{c}{\textsc{SimAnneal~\cite{ng2020siqad}}} & & \multicolumn{2}{c}{\textsc{\emph{QuickSim}}}  \\
					\cmidrule{1-3} \cmidrule{5-5} \cmidrule{7-11} \cmidrule{13-14}				\multicolumn{1}{c}{\multirow{2}{*}{Name}} & \multicolumn{1}{c}{\multirow{2}{*}{\#SiDBs}} & \multicolumn{1}{c}{\multirow{2}{*}{\#inst.}}  &  &  \multicolumn{1}{c}{\multirow{2}{*}{TTS}} & & \multicolumn{2}{c}{\multirow{1}{*}{default}} & &  \multicolumn{2}{c}{\multirow{1}{*}{optimized}} && \multicolumn{2}{c}{\multirow{1}{*}{default}}  \\
					\cmidrule{7-8} \cmidrule{10-11} \cmidrule{13-14}
					\multicolumn{6}{c}{~} &  \multicolumn{1}{c}{acc.} & \multicolumn{1}{c}{TTS} &&   \multicolumn{1}{c}{acc.} &  \multicolumn{1}{c}{TTS} &&  \multicolumn{1}{c}{acc.} & \multicolumn{1}{c}{TTS} \\
					\midrule
					\multirow{1}{*}{Double Wire}   
					& $ 28-30 $&  5& & $1149.21$ & & $2.3$ & $35.80$ & & $1.9$  & $4.87$ && $65.2$ & $0.30$ \\ \addlinespace[0.5em]
					\multirow{1}{*}{CX}   
					& $27-29 $ & 5& & $618.88$ & & $17.5$ & $21.00$ & & $14.5$  & $13.07$ && $64.0$ & $0.25$	\\ \addlinespace[0.5em]
					\multirow{1}{*}{HA}   
					& $ 24-26 $ & 5& & $54.82$ & & $2.8$ & $4.27$ & & $2.2$  & $0.38$ && $96.3$ & $0.04$	\\ \addlinespace[0.5em]
					\multirow{1}{*}{AND}   
					& $21-25$ &5& & $24.31$ & & $31.1$ & $2.33$ & & $20.2$  & $1.49$ && $90.5$ & $0.03$
					\\ \addlinespace[0.5em]
					\multirow{1}{*}{XOR}       
					& $21-23 $ & 5& & $5.65$ & & $38.0$ & $2.13$ & & $24.2$  & $0.42$ && $96.0$ & $0.02$	\\ \addlinespace[0.5em]
					\multirow{1}{*}{OR}   
					& $ 21-23$ & 5& & $5.57$ & & $61.4$ & $0.71$ & & $41.8$  & $0.14$&& $97.7$ & $0.01$	\\ \addlinespace[0.5em]
					\multirow{1}{*}{XNOR}   
					& $21-23 $ & 5& & $5.56$ & & $36.0$ & $1.99$ & & $34.3$  & $0.13$ && $90.7$ & $0.03$\\ \addlinespace[0.5em]
					\multirow{1}{*}{FO2}   
					& $ 20-21$ & 3& & $1.28$ & & $67.2$ & $0.26$ & & $48.2$  & $0.05$ && $98.9$ & $0.01$\\ \addlinespace[0.5em]
					\multirow{1}{*}{NOR}             
					& $19-21$ & 5& & $1.20$ & & $40.2$ & $1.35$ & & $23.0$  & $0.81$ && $95.0$ & $0.02$	\\ \addlinespace[0.5em]
					\multirow{1}{*}{NAND}   
					& $19-21 $ & 5& & $1.19$ & & $56.9$ & $0.67$ &&  $38.3$  & $0.13$ && $99.7$ & $0.01$	\\ \addlinespace[0.5em]
					\multirow{1}{*}{INV}   
					& $ 18-19$ & 6& & $0.53$ & & $78.1$ & $0.35$ & & $52.7$  & $0.09$ && $83.6$ & $0.01$	\\  \addlinespace[0.5em]
					\multirow{1}{*}{Wire}   
					& $ 15-17$ & 6& & $0.08$ & & $84.0$ & $0.24$ & & $39.0$  & $0.13$ && $86.9$ & $0.00$ \\  \midrule
					\multirow{1}{*}{\emph{Mean}}
					&  & & & &&$43.0$ &  &&$28.4$& &&$\mathbf{88.7}$&  \\ 
					\addlinespace[0.5em]
					\multirow{1}{*}{\emph{Total}}
					&  & & &$1868.28$ &&& $71.10$ &&& $21.71$  &&& $\mathbf{0.73}$  \\ 
					\bottomrule
				\end{tabular}
			\end{adjustbox}
		\end{minipage}
	\end{table}
	
	\subsection{Discussion} \label{sec:eval:discussion}
	All obtained results confirm the discussions from \refsec{trade_off}: ExGS obtains the ground state simulation result in a single run, but also requires the largest runtime. SimAnneal is faster but, even after a tedious tuning of the number of annealing cycles for SimAnneal to minimize TTS, a substantial amount of iterations are still required and accuracy is far from satisfactory. Overall: ExGS has high TTS and SimAnneal low accuracy. 

	In contrast, \emph{QuickSim} achieves simulation performances that excel in both accuracy and TTS---outperforming ExGS and SimAnneal on all three evaluations. 
	While, in comparison to SimAnneal, \emph{QuickSim's} simulation accuracy is increased by more than a factor of three, its TTS is improved by more than one order of magnitude at the same time because its increased accuracy does not negatively impact its runtime. Compared to ExGS, its TTS is reduced by several orders of magnitude, while the simulation accuracy is only slightly lower, again emphasizing the overcoming of the \mbox{trade-off} between efficiency and accuracy.
	Thus, a superior ratio between accuracy and single runtime compared to \mbox{state-of-the-art} approaches is achieved by \emph{QuickSim}.
	
	
	\section{Conclusions} \label{sec:concl}
	
	The recent advancement in the design automation and fabrication capabilities of \mbox{\emph{Silicon Dangling Bonds} (SiDBs)} as an emerging computational \mbox{beyond-CMOS} technology was fueled by scientific and commercial interest. In order to keep pace with the progress made in this area, physical simulation is a key tool as it poses the foundation of any design validation workflow.
	
	In this work, the novel algorithm \emph{QuickSim} for physical simulation of SiDB layouts was proposed that outperforms the current \mbox{state-of-the-art} algorithm SimAnneal by more than one order of magnitude and more than a factor of three in terms of \mbox{time-to-solution (TTS)} and accuracy, respectively. 
	
	These results were revealed by an exhaustive experimental evaluation on three different sets of SiDB layouts: \begin{enumerate*}
		\item randomly generated, 
		\item machine learning-generated, and
		\item established ones from the literature,
	\end{enumerate*}
	which constitutes a total of $436$ test cases. 
	
	
	In an effort to support open research and open data, we made the algorithm's implementation, all evaluation data, and the test scripts publicly available. With this approach, we enable the scientific community to build upon our findings and encourage \mbox{follow-up} studies.  
	
	
	

	\bibliographystyle{IEEEtran}
	\bibliography{./bib/IEEEabrv, ./bib/Bibliography.bib}
	
\end{document}